\documentclass[a4paper,12pt]{article} \usepackage{graphics,natbib,times,enumerate,setspace} \usepackage[pagewise,mathlines,displaymath]{lineno} \topmargin=0mm  \oddsidemargin=5mm  \textheight=240mm  \textwidth=150mm 
\usepackage{amsmath,graphicx}
\usepackage[english]{babel}

\newcommand{\nocomma}{}
\newcommand{\tmmathbf}[1]{\ensuremath{\boldsymbol{#1}}}
\newcommand{\tmop}[1]{\ensuremath{\operatorname{#1}}}
\begin{document}

\title{Metric dynamics for membrane transformation through regulated cell
proliferation}
\author{}
\maketitle

\begin{center}
  Hiroshi C. Ito
\end{center}

Department of Evolutionary Studies of Biosystems, SOKENDAI (The Graduate
University for Advanced Studies, Hayama, Kanagawa 240-0193, Japan

Email:hiroshibeetle@gmail.com

\section*{Abstract}

This study develops an equation for describing three-dimensional membrane
transformation through proliferation of its component cells regulated by
morphogen density distributions on the membrane. The equation is developed in
a two-dimensional coordinate system mapped on the membrane, referred to as the
membrane coordinates. When the membrane expands, the membrane coordinates
expand in the same manner so that the membrane is invariant in the
coordinates. In the membrane coordinate system, the transformation of membrane
is described with a time-derivative equation for metric tensors. By defining
relationships between morphogen density distributions and the direction and
rate of cell division, trajectories of membrane transformation are obtained in
terms of the morphogen distributions. An example of the membrane
transformation is shown numerically.

\section{Introduction}

Arthropods have diverse morphologies \citep{regier10:arthropods}. Those
morphologies are made by molting; new epidermis can have significantly
different shape from the old ones through morphogen-regulated proliferation
and/or apotosis of the component cells \citep{warren13:develop}. While
understandings of gene networks for the regulation have been accumulating
rapidly, it is still not clear how the three-dimensional morphologies of
epidermis are generated by morphogen distributions on it. For efficient
investigation on the mechanism underlying this process, this study develops an
equation for metric dynamics describing three-dimensional transformation of a
membrane (corresponding to the new epidermis) through proliferation of its
component cells regulated by morphogen density distributions on it. A unit
element in this equation, referred to as the metric-dynamics equation, is a
(sufficiently) small domain of a membrane, which still has many cells. In this
sense, numerical calculation of the metric-dynamics equation may be more
efficient than the approaches explicitly describing cell divisions.

This paper is structured as follows. Section 2 derives the metric-dynamics
equation. Section 3 numerically shows development of a horn-like structure, as
an application example. Section 4 discusses strong and weak points of the
equation.

\section{Derivations of metric-dynamics equation}

We assume three-dimensional Euclidean coordinate system $\mathbf{R}= ( X, Y,
Z)^T$ and a membrane sheet expressed as a two-dimensional flat plane $Z = 0$.
On the plane, we map a two-dimensional coordinate system $\mathbf{r}= ( x,
y)^T$. When the membrane sheet shrinks, expands and/or be bent, this
coordinate system also changes in the same manner, so that in this membrane
coordinates the membrane sheet is invariant. Thus, any point in the membrane
coordinate system does not change its position through the dynamics. In other
words, in this coordinate system cell division is not followed by growth of
the divided cells.

\subsection{Metric tensor based on cellular distance}

The membrane is expressed as \ $\mathbf{R} ( \mathbf{r}) = ( X ( \mathbf{r}),
Y ( \mathbf{r}), Z ( \mathbf{r}))^T$. We assume that $\mathbf{R} ( \mathbf{r})
= ( x, y, 0)^T$ holds at the initial state. As cell proliferation proceeds, $(
X ( \mathbf{r}), Y ( \mathbf{r}))^T$ deviates from $( x, y)^T$, and \ $Z (
\mathbf{r})$ deviates form the flat plane $Z ( \mathbf{r}) = 0$.

We assume that in coordinate system $\mathbf{R}$ the average cell of any
small (but still sufficiently large for having many cells) region on the
membrane is isotropic and its diameter is equal to a positive constant
$\sigma$ ($\ll 1$), i.e., isometric embedding. Then, the number $\Delta l$ of
cells penetrated by the line segment between $\mathbf{R}$ and $\mathbf{R}+
\Delta \mathbf{R}$ on the membrane is expressed as $| \Delta \mathbf{R} | /
\sigma$ with an additional isotropic-scaling of axes. Then a cellular distance
$\sigma \Delta l$ can be expressed with the corresponding two points
$\mathbf{r}$ and $\mathbf{r}+ \Delta \mathbf{r}$ as
\begin{eqnarray}
  \sigma^2 \Delta l^2 & = & a \Delta x^2 + 2 a b \Delta x \Delta y + c \Delta
  y^2 \nonumber\\
  & = & \left(\begin{array}{c}
    \Delta x\\
    \Delta y
  \end{array}\right)^T \left(\begin{array}{cc}
    a & b\\
    b & c
  \end{array}\right) \left(\begin{array}{c}
    \Delta x\\
    \Delta y
  \end{array}\right) \nonumber\\
  & = & \Delta \mathbf{r}^T \mathbf{G} \Delta \mathbf{r} \nonumber\\
  & = & \Delta \mathbf{r}^T \mathbf{W}^T \mathbf{W} \Delta \mathbf{r} 
\end{eqnarray}
by choosing appropriate coefficients $a, b$, and $c$, where a symmetric matrix
$\mathbf{G}=\mathbf{W}^T \mathbf{W}$ is the metric tensor at $\mathbf{r}$, and
$\mathbf{W}$ is its Cholesky decomposition.

With the metric, we introduce a two-dimensional locally geodesic coordinates
$\tilde{\mathbf{r}}$ around an arbitrary point $\hat{\mathbf{r}}$,
\begin{eqnarray}
  \tilde{\mathbf{r}} - \hat{\mathbf{r}} & = & \mathbf{W} \left[ ( \mathbf{r}-
  \hat{\mathbf{r}}) + \frac{1}{2} \left(\begin{array}{c}
    ( \mathbf{r}- \hat{\mathbf{r}})^T \tmmathbf{\Gamma}^x ( \mathbf{r}-
    \hat{\mathbf{r}})\\
    ( \mathbf{r}- \hat{\mathbf{r}})^T \tmmathbf{\Gamma}^y ( \mathbf{r}-
    \hat{\mathbf{r}})
  \end{array}\right) \right], \nonumber\\
  \mathbf{r}- \hat{\mathbf{r}} & = & \mathbf{W}^{- 1} ( \tilde{\mathbf{r}} -
  \hat{\mathbf{r}}) - \frac{1}{2} \left(\begin{array}{c}
    {}[ \mathbf{W}^{- 1} ( \tilde{\mathbf{r}} - \hat{\mathbf{r}})]^T
    \tmmathbf{\Gamma}^x [ \mathbf{W}^{- 1} ( \tilde{\mathbf{r}} -
    \hat{\mathbf{r}})] + \ldots\\
    {}[ \mathbf{W}^{- 1} ( \tilde{\mathbf{r}} - \hat{\mathbf{r}})]^T
    \tmmathbf{\Gamma}^y [ \mathbf{W}^{- 1} ( \tilde{\mathbf{r}} -
    \hat{\mathbf{r}})] + \ldots
  \end{array}\right), 
\end{eqnarray}
with $\tmmathbf{\Gamma}^x$ and $\tmmathbf{\Gamma}^y$ that contain Christoffel
symbols of the second kind,
\begin{eqnarray}
  \tmmathbf{\Gamma}^x & = & \left(\begin{array}{cc}
    \Gamma^x_{x x} & \Gamma^x_{x y}\\
    \Gamma^x_{x y} & \Gamma^x_{y y}
  \end{array}\right), \nonumber\\
  \tmmathbf{\Gamma}^y & = & \left(\begin{array}{cc}
    \Gamma^y_{x x} & \Gamma^y_{x y}\\
    \Gamma^x_{x y} & \Gamma^x_{y y}
  \end{array}\right), \nonumber\\
  \Gamma^k_{j i} & = & \sum_m \frac{1}{2} G^{- 1}_{k m} \left( \frac{\partial
  G_{i m}}{\partial x_j} + \frac{\partial G_{j m}}{\partial x_i} -
  \frac{\partial G_{j i}}{\partial x_m} \right), 
\end{eqnarray}
where $G^{- 1}_{k m}$ is $( k, m)$ component of $\mathbf{G}^{- 1}$ and $G_{i
m}$ is $( i, m)$ component of $\mathbf{G}$.

Then in coordinate system $\tilde{\mathbf{r}}$ the average cell is isotropic
with constant diameter in the neighborhood of $\hat{\mathbf{r}}$, where
$\sigma^2 \Delta l^2 = | \Delta \tilde{\mathbf{r}} |^2$ holds. From Eqs. (2)
we see at an arbitrary position $\hat{\mathbf{r}}$
\begin{eqnarray}
  \nabla_{\tilde{\mathbf{r}}} \mathbf{r} & = & \tilde{\nabla}
  \mathbf{r}=\mathbf{W}^{- 1}, \nonumber\\
  \nabla_{\mathbf{r}} \tilde{\mathbf{r}} & = & \nabla \tilde{\mathbf{r}}
  =\mathbf{W}, 
\end{eqnarray}
where $\tilde{\nabla}$ and $\nabla$ mean $\nabla_{\tilde{\mathbf{r}}}$ and
$\nabla_{\mathbf{r}}$, respectively.

\subsection{Membrane transformation induced by morphogen distribution}

We assume two types of morphogens: $\alpha$ and $\eta$. Their density
distributions on the membrane are described in $\mathbf{r}$, as $\alpha (
\mathbf{r})$ and $\eta ( \mathbf{r})$. Morphogens $\alpha$ and $\eta$ control
directed and non-directed cell division, respectively. For $\alpha$, we assume
that directed cell division occurs in the direction of gradient of $\alpha$ in
coordinate system $\tilde{\mathbf{r}}$, denoted by $\tilde{\nabla}
\tilde{\alpha}$, with its rate proportional to $b ( \tilde{\nabla}
\tilde{\alpha})$. For $\eta$, the rate of non-directed division is assumed to
be proportional to $\eta ( \mathbf{r})$.

Then from time $t$ to $t' = t + \Delta t$, $\Delta \tilde{\mathbf{r}}$ changes
into $\Delta \tilde{\mathbf{r}}'$, satisfying
\begin{eqnarray}
  \Delta \tilde{\mathbf{r}}' & = & \left[ \mathbf{I}+ \left( c_{\alpha} b ( |
  \tilde{\nabla} \tilde{\alpha} |) \frac{\tilde{\nabla} \tilde{\alpha}
  \tilde{\nabla} \tilde{\alpha}^T}{| \tilde{\nabla} \tilde{\alpha} |^2} +
  c_{\eta} \eta \mathbf{I} \right) \Delta t \right] \Delta \tilde{\mathbf{r}}
  \\
  & = & [ \mathbf{I}+\mathbf{A} \Delta t] \Delta \tilde{\mathbf{r}} \nonumber
\end{eqnarray}
with proportionality coefficients $c_{\alpha}$ and $c_{\eta}$, where $(
\mathbf{r})$ is omitted. Without loss of generality, we assume that $\alpha (
\mathbf{r})$ and $\eta ( \mathbf{r})$ are scaled so that $c_{\alpha} =
c_{\eta} = 1$. The cellular distance between $\mathbf{r}$ and $\mathbf{r}+
\Delta \mathbf{r}$ at $t' = t + \Delta t$ is given by
\begin{eqnarray}
  \sigma^2 \Delta l'^2 & = & \Delta \tilde{\mathbf{r}}'^T \Delta
  \tilde{\mathbf{r}}' = \Delta \tilde{\mathbf{r}}^T [ \mathbf{I}+\mathbf{A}
  \Delta t]^T [ \mathbf{I}+\mathbf{A} \Delta t] \Delta \tilde{\mathbf{r}} . 
\end{eqnarray}

\subsection{Metric dynamics in membrane coordinate system}

To express Eq. (6) in the membrane coordinate system, we express density
distribution of $\alpha$ in coordinate system $\tilde{\mathbf{r}}$ as
\begin{eqnarray}
  \tilde{\alpha} ( \tilde{\mathbf{r}}) & = & \frac{\alpha ( \mathbf{r})}{\|
  \mathbf{W} \|} = \frac{\alpha ( \mathbf{W}^{- 1} \tilde{\mathbf{r}} +
  \ldots)}{\| \mathbf{W} \|}, 
\end{eqnarray}
which gives a relationship
\begin{eqnarray}
  \tilde{\nabla} \tilde{\alpha} = \nabla_{\tilde{\mathbf{r}}} \tilde{\alpha} &
  = & \nabla_{\tilde{\mathbf{r}}} \frac{\alpha ( \mathbf{W}^{- 1}
  \tilde{\mathbf{r}} + \ldots)}{\| \mathbf{W} \|} = \frac{1}{\| \mathbf{W} \|}
  \nabla_{\tilde{\mathbf{r}}} \mathbf{r}^T \nabla_{\mathbf{r}} \alpha =
  \frac{1}{\| \mathbf{W} \|} \mathbf{W}^{- 1 T} \nabla \alpha . 
\end{eqnarray}
Here we assume that
\begin{eqnarray}
  b ( | \tilde{\nabla} \tilde{\alpha} |) & = & \| \mathbf{G} \| |
  \tilde{\nabla} \tilde{\alpha} |^2, 
\end{eqnarray}
which gives a simple result as follows. Substituting Eqs. (8) and (9) into Eq.
(5) gives
\begin{eqnarray}
  \mathbf{A} & = & \mathbf{W}^{- 1 T} \nabla \alpha \nabla \alpha^T
  \mathbf{W}^{- 1} + \eta \mathbf{I}, 
\end{eqnarray}
which upon substitution into Eq. (6) gives
\begin{eqnarray}
  \sigma^2 \Delta l'^2 & = & \Delta \mathbf{r}^T \mathbf{W}^T [ ( 1 + 2 \eta
  \Delta t) \mathbf{I}+ 2\mathbf{W}^{- 1 T} \nabla \alpha \nabla \alpha^T
  \mathbf{W}^{- 1} \Delta t + O ( \Delta t^2)] \mathbf{W} \Delta \mathbf{r}
  \nonumber\\
  & = & \Delta \mathbf{r}^T [ ( 1 + 2 \eta \Delta t) \mathbf{G}+ 2 \nabla
  \alpha \nabla \alpha^T \Delta t + O ( \Delta t^2)] \Delta \mathbf{r}
  \nonumber\\
  & = & \Delta \mathbf{r}^T [ \mathbf{G}+ 2 ( \eta \mathbf{G}+ \nabla \alpha
  \nabla \alpha^T) \Delta t + O ( \Delta t^2)] \Delta \mathbf{r} \nonumber\\
  & = & \Delta \mathbf{r}^T \mathbf{G}' \Delta \mathbf{r}. 
\end{eqnarray}
Therefore, the time derivative of metric $\mathbf{G}$ is obtained as
\begin{eqnarray}
  \frac{d\mathbf{G}}{d t} & = & \lim_{\Delta t \rightarrow 0}
  \frac{\mathbf{G}' -\mathbf{G}}{\Delta t} = 2 \eta \mathbf{G}+ 2 \nabla
  \alpha \nabla \alpha^T . 
\end{eqnarray}
From this equation, the time derivative of linear cell density $q (
\mathbf{e}) = \sqrt{\mathbf{e}^T \mathbf{G}\mathbf{e}}$, which is the
multiplication of $\sigma$ by the number of cells penetrated by unit vector
$\mathbf{e}$ in coordinate system $\mathbf{r}$, is given by
\begin{eqnarray}
  \frac{d q ( \mathbf{e})^2}{d t} & = & \mathbf{e}^T \frac{d\mathbf{G}}{d t}
  \mathbf{e}= 2\mathbf{e}^T [ \eta \mathbf{G}+ \nabla \alpha \nabla \alpha^T]
  \mathbf{e}, \\
  \frac{d q ( \mathbf{e})}{d t} & = & \frac{1}{q ( \mathbf{e})} \mathbf{e}^T [
  \eta \mathbf{G}+ \nabla \alpha \nabla \alpha^T] \mathbf{e}. 
\end{eqnarray}

\subsection{Solution of metric-dynamics equation}

Solving Eq. (12) gives the metric and linear cell density after cell
proliferation from time $t = 0$ to an arbitrary $t = \tau$,
\begin{eqnarray}
  \mathbf{G} ( \tau) & = & e^{2 \eta \tau} \left[ \int_0^{\tau} 2 \nabla
  \alpha \nabla \alpha^T e^{- 2 \eta t} \tmop{dt} +\mathbf{G}_0 \right], \\
  q^2 ( \mathbf{e}, \tau) & = & \mathbf{e}^T \mathbf{G} ( \tau) \mathbf{e}. 
\end{eqnarray}
with $\mathbf{G}_0 =\mathbf{G} ( 0)$. If $\eta$ and $\nabla \alpha$ is
constant along time, Eq. (15) is simplified into
\begin{eqnarray}
  \mathbf{G} ( \tau) & = & \left\{ \begin{array}{lll}
    \left[ \mathbf{G}_0 + \frac{\nabla \alpha \nabla \alpha^T}{\eta} \right]
    \exp ( 2 \eta \tau) - \frac{\nabla \alpha \nabla \alpha^T}{\eta} &
    \tmop{for} & \eta \neq 0\\
    \mathbf{G}_0 + 2 \tau \nabla \alpha \nabla \alpha^T & \tmop{for} & \eta =
    0.
  \end{array} \right. 
\end{eqnarray}

To see the shape of the membrane given by Eq. (17), we consider that $\alpha (
\mathbf{r})$ is expressed in coordinate system $\mathbf{R}$ as a
two-dimensional surface $\mathbf{R}_{\alpha} = ( x, y, \alpha (
\mathbf{r}))^T$. Then distance between the two points on the surface,
$\mathbf{R}_{\alpha}$ and $\mathbf{R}_{\alpha} + \Delta \mathbf{R}_{\alpha}$
(corresponding to $\mathbf{r}$ and $\mathbf{r}+ \Delta \mathbf{r}$), can be
expressed as
\begin{eqnarray}
  \sigma^2 \Delta l_{\alpha}^2 & = & \Delta x^2 + \Delta y^2 + | \nabla
  \alpha^T \Delta \mathbf{r} |^2, \nonumber\\
  & = & \Delta \mathbf{r}^T \left(\begin{array}{cc}
    1 & 0\\
    0 & 1
  \end{array}\right) \Delta \mathbf{r}+ \Delta \mathbf{r}^T \nabla \alpha
  \nabla \alpha^T \Delta \mathbf{r} \nonumber\\
  & = & \Delta \mathbf{r}^T [ \mathbf{I}+ \nabla \alpha \nabla \alpha^T]
  \Delta \mathbf{r}, 
\end{eqnarray}
which gives
\begin{eqnarray}
  \mathbf{G}_{\alpha} & = & \mathbf{I}+ \nabla \alpha \nabla \alpha^T . 
\end{eqnarray}
Thus, the membrane may have a similar shape to that of $\mathbf{R}_{\alpha}$.
Especially, $\mathbf{G}_0 = 2 \tau \mathbf{I}$ gives $\mathbf{G} ( \tau)
=\mathbf{G}_{\alpha}$. Even for \ $\mathbf{G}_0 \neq 2 c_{\alpha} \tau
\mathbf{I}$, a sufficiently large $| \nabla \alpha |$ allows $\mathbf{G} (
\tau) \approx 2 c_{\alpha} \tau \mathbf{G}_{\alpha}$.

\subsection{Modification by other morphogens}

Here we assume that distributions of $\alpha$ and $\eta$ are constant along
time in the membrane coordinate system. In order to modify the basic membrane
structure formed by morphogens $\alpha$ and $\eta$, we introduce additional
morphogens $\beta$ and $\theta$. Morphogens $\beta$ and $\theta$ curve and
twist the basic structure, respectively, as explained below.

First, $\beta$ accelerates the rate of directed cell division in one side of
the basic structure, and suppresses that in the other side, which can be
realized by introducing an increasing function of $[ \tilde{\nabla}
\tilde{\alpha}^T \tilde{\nabla} \tilde{\beta}]_{t = 0} = \| \mathbf{G}_0 \|^{-
1} \nabla \alpha^T \mathbf{G}_0^{- 1} \nabla \beta$, denoted by $g ( [
\tilde{\nabla} \tilde{\alpha}^T \tilde{\nabla} \tilde{\beta}]_{t = 0})$, \
into Eq. (5) as
\begin{eqnarray}
  \Delta \tilde{\mathbf{r}}' & = & [ \mathbf{I}+ ( g ( [ \tilde{\nabla}
  \tilde{\alpha}^T \tilde{\nabla} \tilde{\beta}]_{t = 0}) \tilde{\nabla}
  \tilde{\alpha} \tilde{\nabla} \tilde{\alpha}^T + \eta \mathbf{I}) \Delta t]
  \Delta \tilde{\mathbf{r}} . 
\end{eqnarray}
For simplicity, we choose $g ( [ \tilde{\nabla} \tilde{\alpha}^T
\tilde{\nabla} \tilde{\beta}]_{t = 0}) = \exp ( [ \tilde{\nabla}
\tilde{\alpha}^T \tilde{\nabla} \tilde{\beta}]_{t = 0})^2$. (Curving can also
be realized by acceleration/suppression of non-directed cell division, by
multiplying $\eta \mathbf{I}$ by $[ \tilde{\nabla} \tilde{\alpha}^T
\tilde{\nabla} \tilde{\beta}]_{t = 0}$.)

Second, $\theta$ rotates the direction of directed cell division by $\theta$
in the initial geodesic coordinate system (i.e., rotation of $\tilde{\nabla}
\tilde{\alpha} = \| \mathbf{W} \|^{- 1} \mathbf{W}^{- 1 T} \nabla \alpha$ by
$\theta$ in $\tilde{\mathbf{r}}$ at $t = 0$), which further transforms Eq.
(20) into
\begin{eqnarray}
  \Delta \tilde{\mathbf{r}}' & = & [ \mathbf{I}+ O ( \Delta t^2)] \Delta
  \tilde{\mathbf{r}} \\
  &  & + ( \exp ( [ \tilde{\nabla} \tilde{\alpha}^T \tilde{\nabla}
  \tilde{\beta}]_{t = 0})^2 [ \mathbf{W}^{- 1 T} \tilde{\tmmathbf{\Theta}}
  \nabla \alpha] [ \mathbf{W}^{- 1 T} \tilde{\tmmathbf{\Theta}} \nabla
  \alpha]^T + \eta \mathbf{I}) \Delta t \Delta \tilde{\mathbf{r}} \nonumber\\
  & = & [ \mathbf{W}+ ( \exp ( \| \mathbf{G}_0 \|^{- 1} \nabla \alpha^T
  \mathbf{G}_0^{- 1} \nabla \beta)^2 \mathbf{W}^{- 1 T}
  \tilde{\tmmathbf{\Theta}} \nabla \alpha \nabla \alpha^T
  \tilde{\tmmathbf{\Theta}}^T + \eta \mathbf{W}) \Delta t + O ( \Delta t^2)]
  \Delta \mathbf{r} \nonumber\\
  & = & [ \mathbf{W}+ ( \mathbf{W}^{- 1 T} \mathbf{a}\mathbf{a}^T + \eta
  \mathbf{W}) \Delta t + O ( \Delta t^2)] \Delta \mathbf{r}, \nonumber
\end{eqnarray}
where
\begin{eqnarray}
  \mathbf{a} & = & \exp ( \| \mathbf{G}_0 \|^{- 1} \nabla \alpha^T
  \mathbf{G}_0^{- 1} \nabla \beta) \tilde{\tmmathbf{\Theta}} \nabla \alpha,
  \nonumber\\
  \tilde{\tmmathbf{\Theta}} & = & \mathbf{W}_0^T
  \tmmathbf{\Theta}\mathbf{W}_0^{- 1 T}, \nonumber\\
  \tmmathbf{\Theta} & = & \left(\begin{array}{cc}
    \cos \theta & \sin \theta\\
    - \sin \theta & \cos \theta
  \end{array}\right), 
\end{eqnarray}
and $\mathbf{W}_0^T \mathbf{W}_0 =\mathbf{G}_0$. Then we obtain,
\begin{eqnarray}
  \sigma^2 \Delta l'^2 & = & \Delta \mathbf{r}^T [ \mathbf{G}+ 2 ( (
  \mathbf{a}\mathbf{a}^T + \eta \mathbf{G}) \Delta t) + O ( \Delta t^2)]
  \Delta \mathbf{r}, \\
  \frac{d\mathbf{G}}{d t} & = & 2\mathbf{a}\mathbf{a}^T + 2 \eta \mathbf{G},
  \\
  \mathbf{G} ( \tau) & = & \left\{ \begin{array}{lll}
    \left[ \mathbf{G}_0 + \frac{\mathbf{a}\mathbf{a}^T}{\eta} \right] \exp ( 2
    \eta \tau) - \frac{\mathbf{a}\mathbf{a}^T}{\eta} & \tmop{for} & \eta \neq
    0\\
    \mathbf{G}_0 + 2 \tau \mathbf{a}\mathbf{a}^T & \tmop{for} & \eta = 0.
  \end{array} \right. 
\end{eqnarray}

The above formulation uses $\tilde{\nabla} \tilde{\alpha}$ and $\tilde{\nabla}
\tilde{\beta}$ at $t = 0$ for the modification. We can also use
$\tilde{\nabla} \tilde{\alpha}$ and $\tilde{\nabla} \tilde{\beta}$ at each
time $t$ instead (i.e., removing the subscript `0' from $\mathbf{G}_0$ and
$\mathbf{W}_0$ in Eqs. (22)), which might be easier to realize in biological
systems. However, this choice gives more complicated equation than Eq. (25).

In the beginning of subsection 2.1 we assume that at $t = 0$ the membrane is
flat, $\mathbf{R} ( \mathbf{r}) = ( X ( \mathbf{r}), Y ( \mathbf{r}), Z (
\mathbf{r}))^T = ( x, y, 0)^T$, which implies $\mathbf{G}_0 =\mathbf{I}$ for
all $\mathbf{r}$. Even if $\mathbf{G}_0$ varies along $\mathbf{r}$, i.e., the
membrane is not flat at the initial state, Eqs. (23-25) are unchanged.

\section{Numerical calculation of membrane transformation}

\begin{figure}[h]
  (a) \ \ \ \ \ \ \ \ \ \ \ \ \ \ \ \ \ \ \ \ \ \ \ \ \ \ \ \ \ \ \ \ \ \ \ \
  \ \ \ \ \ \ \ \ \ \ \ \ \ \ \ \ \ (b) \ \ \ \ \ \ \
  
  \resizebox{6cm}{6cm}{\includegraphics{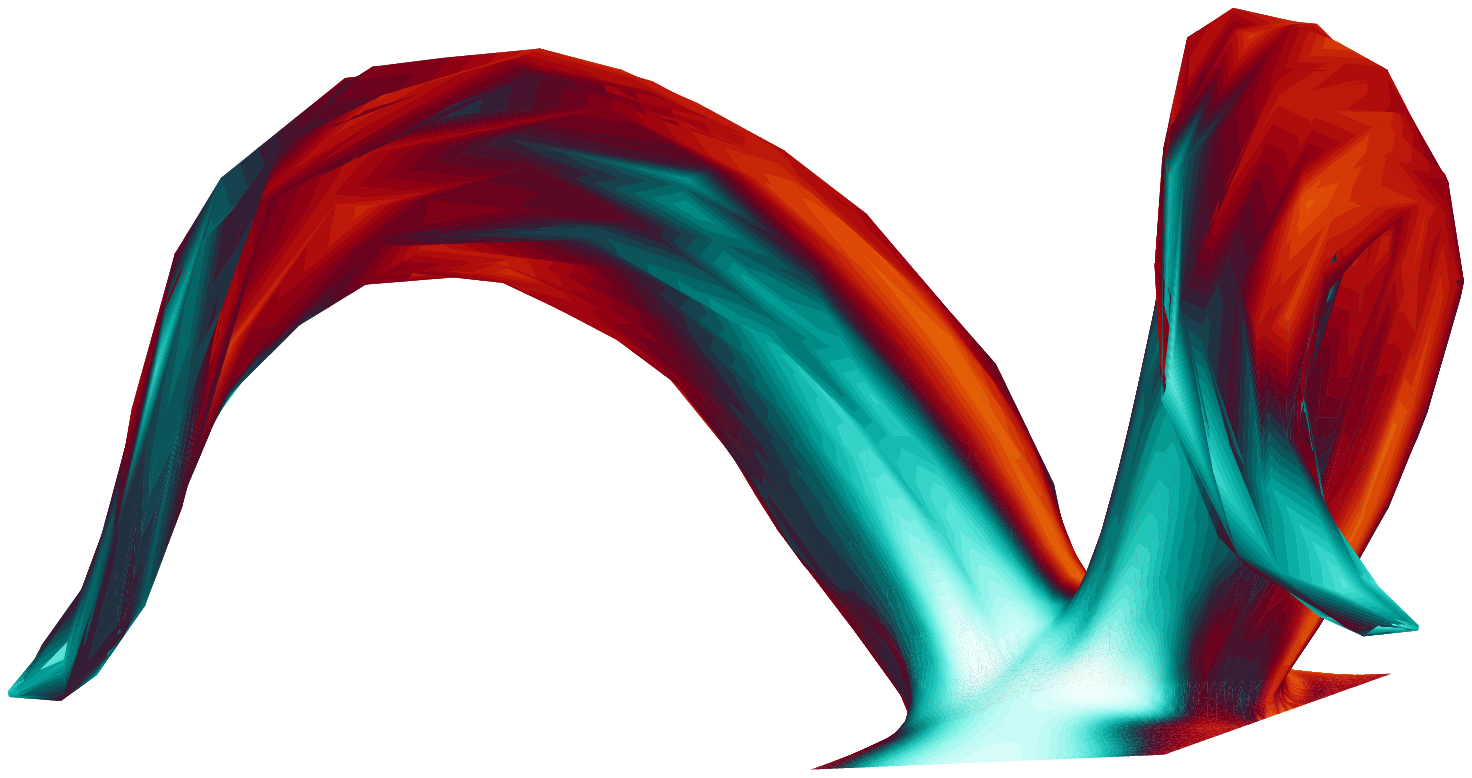}}
  \ \
  \resizebox{6cm}{6cm}{\includegraphics{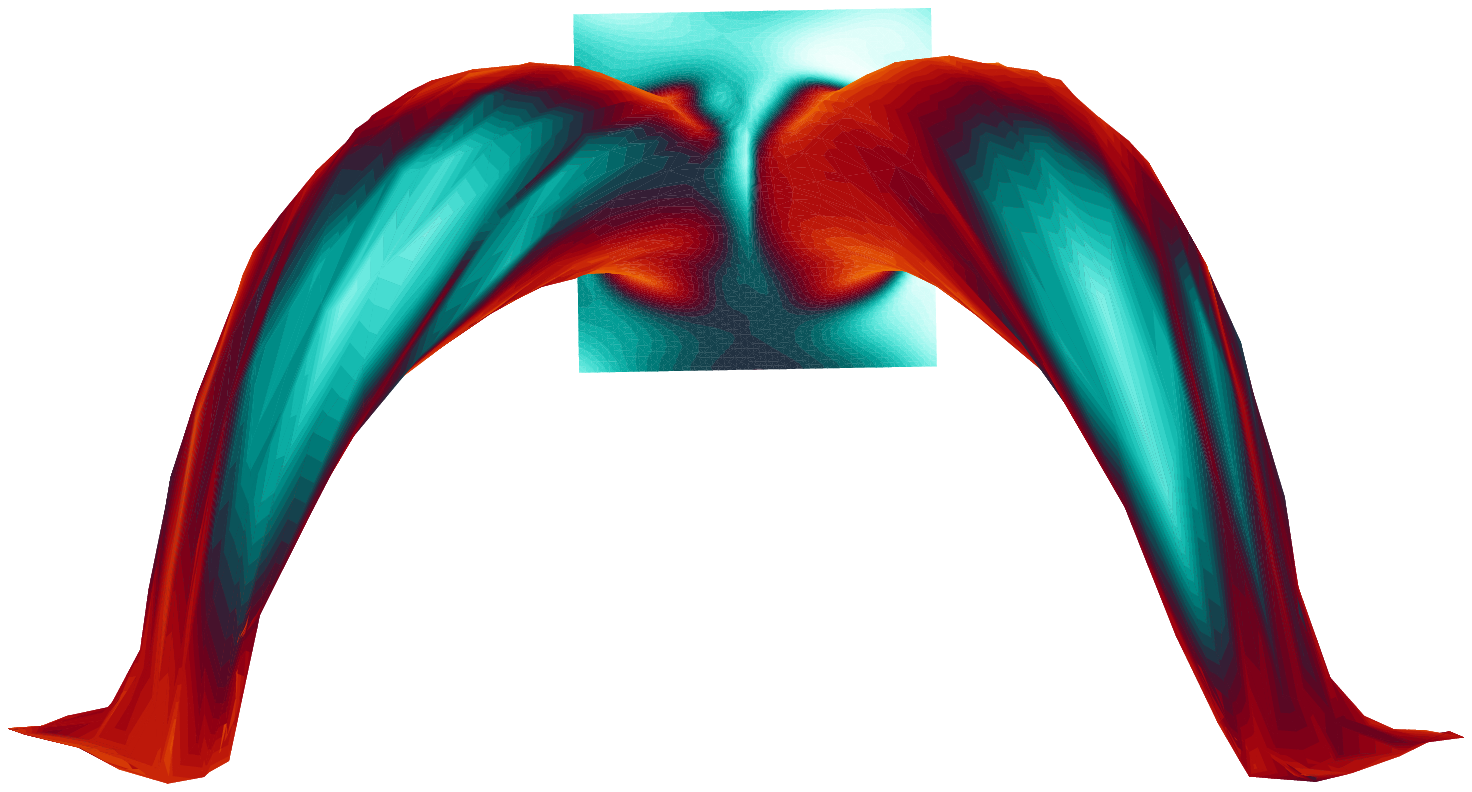}}
  \caption{Transformation of a flat membrane through cell proliferation
  regulated by morphogens. Parameters are $\tau = 0.6$, $d_a = 1,
  \sigma_{\alpha} = 0.5 \nocomma, c_{\alpha} = 30, c_{\beta} = - 0.15$, and
  $c_{\theta} = 0.02$. (a) and (b) show the same membrane seen from different
  perspectives.}
\end{figure}

We show an example of numerical calculation of membrane transformation. For
given density distributions of $\alpha$, $\eta$, $\beta$, $\theta$, and the
initial metric $\mathbf{G}_0$, we can calculate $\mathbf{G} ( \tau)$. The
membrane structure $\mathbf{R} ( \mathbf{r}) = ( X ( \mathbf{r}), Y (
\mathbf{r}), Z ( \mathbf{r}))^T$ at $t = \tau$ can be calculated by solving
\begin{eqnarray}
  \lim_{| \Delta \mathbf{r} | \rightarrow \tmmathbf{0}} \left[ \frac{|
  \mathbf{R} ( \mathbf{r}+ \Delta \mathbf{r}) -\mathbf{R} ( \mathbf{r})
  |^2}{\Delta \mathbf{r}^T \mathbf{G} ( \tau) \Delta \mathbf{r}} \right] & = &
  1 
\end{eqnarray}
with Eq. (25). In this example, for $\alpha$ we assume a distribution
\begin{eqnarray}
  \alpha ( \mathbf{r}) & = & c_{\alpha} \left[ \exp \left( - \frac{|
  \mathbf{r}- ( d_{\alpha}, 0)^T |^2}{2 \sigma_{\alpha}^2} \right) + \exp
  \left( - \frac{| \mathbf{r}- ( - d_{\alpha}, 0)^T |^2}{2 \sigma_{\alpha}^2}
  \right) \right], 
\end{eqnarray}
which have two peaks for a sufficiently large $d_{\alpha}$. For the other
morphogens, we assume
\begin{eqnarray}
  \eta ( \mathbf{r}) & = & 0 \nonumber\\
  \beta ( \mathbf{r}) & = & c_{\beta} y \nonumber\\
  \theta ( \mathbf{r}) & = & \left\{ \begin{array}{lll}
    c_{\theta} & \tmop{for} & x > 0\\
    - c_{\theta} & \tmop{for} & x < 0
  \end{array} \right. 
\end{eqnarray}
For efficiency in solving Eq. (26), we assume a bending elasticity of the
membrane, and a weak water pressure from inside of the membrane (i.e., a
constant outward pressure in parallel with the normal vector at each point on
the membrane). Figure 1 shows a transformed membrane structure for
$\mathbf{G}_0 =\mathbf{W}_0 =\mathbf{I}$ (i.e., the initial membrane is flat),
$\tau = 0.6$, $d_a = 1, \sigma_{\alpha} = 0.5 \nocomma, c_{\alpha} = 30,
c_{\beta} = - 0.15$, and $c_{\theta} = 0.02$.

\section{Discussion}

As the initial metric $\mathbf{G}_0$ can vary along $\mathbf{r}$ for the
metric-dynamics equation, Eq. (25), the membrane transformation can be
repeatedly applied to the same membrane by resetting morphogen distributions
while keeping $\mathbf{G} ( \tau)$ (like as repeated molting of arthropods),
which can generate various complex structures. Our approach may be more
efficient than models describing cell-level dynamics, because the number of
vertices required for describing the membrane may be smaller. On the other
hand, a weak point of our approach is that we have to define relationships
between the local properties of morphogen distributions (e.g., gradients) and
the manners of local membrane extensions. Cell-based approaches can examine
the validity of those relationships and improve them.

\bibliographystyle{/home/itoh9/pop-ecol}

\bibliography{/home/itoh9/itobib}

\begin{thebibliography}{2}
\expandafter\ifx\csname natexlab\endcsname\relax\def\natexlab#1{#1}\fi
\expandafter\ifx\csname url\endcsname\relax
  \def\url#1{\texttt{#1}}\fi
\expandafter\ifx\csname urlprefix\endcsname\relax\def\urlprefix{URL }\fi

\bibitem[{Regier et~al.(2010)Regier, Shultz, Zwick, Hussey, Ball, Wetzer,
  Martin, and Cunningham}]{regier10:arthropods}
Regier~JC, Shultz~JW, Zwick~A, Hussey~A, Ball~B, Wetzer~R, Martin~JW,
  Cunningham~CW (2010) Arthropod relationships revealed by phylogenomic
  analysis of nuclear protein-coding sequences. Nature 463: 1079--1083

\bibitem[{Warren et~al.(2013)Warren, Gotoh, Dworkin, Emlen, and
  Lavine}]{warren13:develop}
Warren~IA, Gotoh~H, Dworkin~IM, Emlen~DJ, Lavine~LC (2013) A general mechanism
  for conditional expression of exaggerated sexually-selected traits. Bioessays
  35: 889--899

\end{thebibliography}

\end{document}